\documentclass[aps,pre,twocolumn,showpacs]{revtex4}  
\usepackage{graphicx}  
\usepackage{dcolumn}   
\usepackage{bm}        
\usepackage{amssymb}   
\usepackage{float}
\usepackage{enumerate}

\begin{document}

\hspace{5.2in}

\title{Impact of symmetry breaking in networks of globally coupled oscillators}
\author{K. Premalatha$^{1}$, V. K. Chandrasekar$^{2}$, M. Senthilvelan$^{1}$, M. Lakshmanan$^{1}$}
\address{$^1$Centre for Nonlinear Dynamics, School of Physics, Bharathidasan University, Tiruchirappalli - 620 024, Tamil Nadu, India.\\
$^2$Centre for Nonlinear Science \& Engineering, School of Electrical \& Electronics Engineering, SASTRA University, Thanjavur -613 401,Tamilnadu, India.}
\begin{abstract}
We analyze the consequences of symmetry breaking in the coupling in a network of globally coupled identical Stuart-Landau oscillators.  We observe that symmetry breaking leads to increased disorderliness in the dynamical behavior of oscillatory states and consequently results in a rich variety of dynamical states.  Depending on the strength of the nonisochronicity parameter, we find various dynamical states such as amplitude chimera, amplitude cluster, frequency chimera and frequency cluster states.  In addition we also find disparate transition routes to recently observed chimera death state in the presence of symmetry breaking even with global coupling. We also analytically verify the chimera death region which corroborates the numerical results.  The above results are compared with that of the symmetry preserving case as well.  
\end{abstract}

\maketitle

\section{Introduction}
\par The study of complex networks has attracted much attention in the fields of physics, chemistry, social sciences, etc. under various coupling topologies \cite{1,2,3,4,5,6}.  For instance collective phenomena in complex networks have been studied profoundly in the past decade which depend on both the peculiarity of the oscillators and the nature of the coupling and its strength.  In particular much of the recent studies in nonlinear dynamics have been devoted to understand the dynamics of globally coupled oscillators due its applicability in real world networks \cite{7,8,9,10,11}. 
\par Out of the various dynamical behaviors in identical nonlinear coupled systems, the existence of hybrid states (chimera states) of combining both synchronized and desynchronized behaviors \cite{12,13,14,15,16,17,18} has received  wide attention because of its connection with many real world applications such as uni-hemispheric sleep \cite{19}, neural networks \cite{20} and so on.  Moreover, chimera states have also been observed in maps \cite{21}, complex networks \cite{22}, time discrete and continuous chaotic systems \cite{23}.  The emergence of chimera states in ensembles of oscillators has been studied theoretically \cite{24,25,26,27,28,29,30,31,32} for more than a decade and experimentally \cite{33,34,35} reported recently under nonlocal interactions.  
\par A few years ago Daido and Nakanishi in Ref. \cite{6} have observed a new interesting phenomenon called swing of synchronized states, while studying the inhomogeneity induced by the introduction of diffusive coupling in globally coupled Stuart-Landau oscillators without symmetry breaking in the coupling.  They found that the synchronized state which has been destabilized because of the increase in the coupling strength is found to be restabilized for further raise of it.  The diffusion in globally coupled systems induces the synchronized state mediated by the so called cluster states.
\par However recent studies suggest that coexistence behavior of chimera states is also observed in globally coupled networks as in the case of systems with nonlocal coupling.   Sethia and Sen have pointed out the emergence of amplitude mediated chimera states (AMC) even in globally coupled oscillators without symmetry breaking in the coupling \cite{36} in a system of Ginzburg-Landau oscillators as the specific example (note that the Stuart-Landau oscillators correspond to a special choice of a control parameter of this system).  A question to ask is whether the amplitude mediated chimera state (AMC is also known as frequency chimera) can exist for global coupling {\it with symmetry breaking}.  We analyze this question in the present work and report the existence of AMC in the case of symmetry breaking in the coupling also, which further leads to a rich variety of dynamical states than the case where the coupling is of symmetry preserving type.      
\par On the other hand,  symmetry breaking instability in a network of coupled oscillators with nonlocal coupling also leads to the existence of stable inhomogeneous steady states.  Interplay of nonlocality with symmetry breaking leads to new a dynamical state, namely chimera death, which was reported by Zakharova et al \cite{37} recently.  In the chimera death state,  the oscillators in the network partition into two coexisting domains, where in one domain neighboring nodes occupy the same branch of the inhomogeneous steady state (spatially coherent oscillation death (OD)) whereas in the other domain neighboring nodes are randomly distributed among the different branches of the inhomogeneous steady state (spatially incoherent OD).  In the present study, we will also investigate whether such chimera death states can exist {\it in the globally coupled system with symmetry breaking as well}.  Indeed we show that such states do exist in this case. 
\par In this article, we are motivated to study the detailed dynamical behavior of the globally coupled Stuart-Landau oscillators in the presence of symmetry breaking in the coupling and compare it with the case when the symmetry is preserved.  We show that the influence of symmetry breaking in the coupling leads to increased disorder in the nature of dynamical states and we observe that their regions also get widened.  In \cite{36}, Sethia and Sen observed that in the case of symmetry preserved coupling the synchronized state is mediated through the frequency chimera state.  In addition to this we observe that the synchronized state is mediated through the amplitude chimera state (different from AMC) as well in both the cases of symmetry breaking and symmetry preserving couplings.  We illustrate the above results with the help of characteristic measures such as standard deviation and strength of incoherence.  Moreover, we present the results for different transition routes to chimera death and also analytically verify the chimera death regions which nearly matches with the numerical results.        
 \par This paper is organized as follows.  In section $II$, we introduce the model of globally coupled Stuart-Landau oscillators with two different couplings that we have considered for our simulation and present the results obtained from the analysis of the various dynamical states.   In section $III$, we give the phase diagrams to illustrate the different transition routes.  We summarize our findings in section $IV$.     
\section{DYNAMICS OF GLOBALLY COUPLED STUART-LANDAU OSCILLATORS IN THE PRESENCE OF SYMMETRY PRESERVING AND SYMMETRY BREAKING COUPLINGS}
\subsection{Model}
In order to exemplify our results, we consider an array of globally coupled identical Stuart-Landau oscillators with two types of couplings:\\
(i) Symmetry preserving coupling: 
\begin{equation}
\dot{w_j}=w_j-(1- ic)|w_j|^2 w_j+\epsilon (\overline{w}-w_j),
\label{g}
\end{equation}
(ii) Symmetry broken coupling:
\begin{equation}
\dot{w_j}=w_j-(1- ic)|w_j|^2 w_j+\epsilon (\overline{Re(w)}-Re(w_j)),
\label{gs}
\end{equation}
where $j=1,2,3,...N$, $N$ being the number of oscillators, $w_j=x_j+iy_j$, $\overline{w}=(1/N)\sum_{j=1}^{N}w_j$, $\overline{Re(w)}=(1/N)\sum_{j=1}^{N}x_j$ is the mean field through the real part of the amplitude,  $\epsilon$ is the coupling constant and $c$ is the nonisochronicity parameter.  In our simulations, we choose the number of oscillators $N$ to be equal to 100 and in order to solve the equations (\ref{g}) and (\ref{gs}), we use the fourth order Runge-Kutta method with time step 0.01.  Note that Eq. (\ref{g}) preserves the gauge symmetry $w_j \rightarrow w_j^\prime=w_je^{i\theta}$,  $\theta \in \mathbb{R}$, or equivalently the rotational symmetry in the ($x,y$) plane, while it is broken in (\ref{gs}). 
\par  In Ref. \cite{6} Daido and Nakanishi have identified the phenomenon of swing in synchronized states in Eq. (\ref{g}) which is mediated through cluster states.  In our study we intensely study this phenomenon and clearly distinguish the dynamical regions in the following sections.  We also compare the results with the system of oscillators with symmetry breaking in the coupling.  It may also be noted that Eq. (\ref{g}) is the special case of the Ginzburg-Landau equation 
\begin{equation}
\dot{w_j}=w_j-(1- ic)|w_j|^2 w_j+\epsilon(1+iC_1) (\overline{w}-w_j)
\label{gz}
\end{equation}
with the parameter chosen as $C_1=0$.  Occurrence of AMC in (\ref{gz}) in the ($C_1,\epsilon$) phase space is discussed in detail in Ref. {\cite{36}} by Sethia and Sen.
\begin{figure}[]
\begin{center}
 \includegraphics[width=1.0\linewidth]{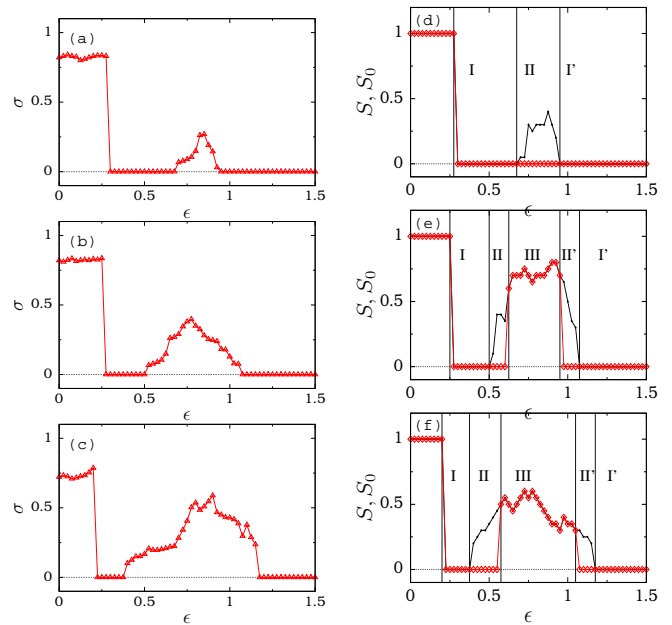}
\end{center}
\caption{Standard deviation ($\sigma$) of the system (\ref{g}) with symmetry preserving coupling for different values of $c$ (a) for $c=3.0$, (b) $c=5.0$, (c) $c=7.0$, and (d-f) show their corresponding strength of incoherence $S$ (red color line) and $S_0$ (black color line).  Regions-I, I$^\prime$ correspond to synchronized states, regions-II, II$^\prime$ show the cluster states and the region-III represents chimera states.}
\label{fg1}
\end{figure} 
\begin{figure}[]
\begin{center}
 \includegraphics[width=1.0\linewidth]{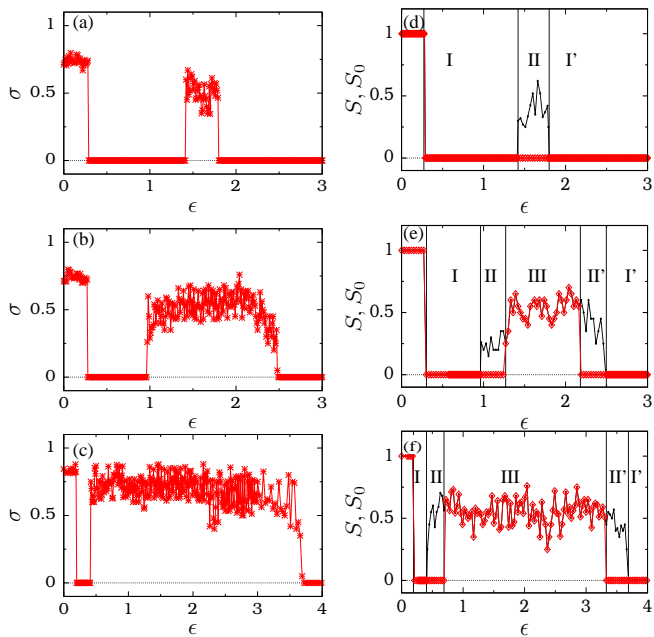}
\end{center}
\caption{Standard deviation ($\sigma$) of the system (\ref{gs}) with symmetry broken coupling for different values of $c$ (a) for $c=2.5$, (b) $c=4.0$, (c) $c=7.0$, and (d-f) show their corresponding strength of incoherence $S$ (red color line) and $S_0$ (black color line).  Regions-I, I$^\prime$ correspond to synchronized states, regions-II, II$^\prime$ show the cluster states and the region-III represents chimera states.}
\label{fg2}
\end{figure} 
\subsection{Quantification of various dynamical states}
 \par We study the characteristic nature of the dynamical states including desynchronized, chimera/cluster and synchronized states in Eqs. (\ref{g}) and (\ref{gs}) with the help of the quantitative measures such as standard deviation \cite{6} and strength of incoherence \cite{39,40} which are explained in the following subsections.
\subsubsection{CHARACTERIZATION WITH RESPECT TO STANDARD DEVIATION}
\par In this section, we use the measure standard deviation for the real part of the amplitudes as used by Daido and Nakanishi \cite{6} defined by 
\begin{equation} 
\sigma=\langle(\overline{\vert x_j-\overline{x_j}\vert^2})^{1/2}\rangle,
\label{sd}
\end{equation}
where the bar denotes average over $1 \le j \le N$ and the angular bracket stands for the time average.  It is clear that the standard deviation of the system that is calculated from (\ref{sd}) is zero for the synchronized state and nonzero for the desynchronized state.
\par Depending on the strength of the nonisochronicity parameter  $c$ and coupling strength $\epsilon$, the system of oscillators attains different dynamical states.  In order to identify the dynamical behavior of these states, to start with we make use of the above mentioned standard deviation measure for various strengths of $c$ and $\epsilon$.  With this intention, we demonstrate the behavior of standard deviation $\sigma$ as a function of $\epsilon$ in Figs. \ref{fg1}(a-c) for symmetry preserved coupling (system (\ref{g})) and in Figs. \ref{fg2}(a-c) for symmetry broken coupling (system (\ref{gs})) for three different choices of $c$. \\
{\bf(a)~~Symmetry preserved global coupling:}\\
In Fig. \ref{fg1}(a), we fix the nonisochronicity parameter at $c=3.0$ and one can observe that $\sigma$ takes nonzero values for $\epsilon< 0.275$ which implies the desynchronization of oscillators in this region.  For the values of $\epsilon$ in the region $0.275< \epsilon < 0.675$ the oscillators are synchronized where $\sigma$ decreases to zero value.  As $\epsilon$ increases $\sigma$ takes again nonzero values in the range $ 0.675 < \epsilon < 0.925$ which indicates desynchronization among the oscillators.  By increasing $\epsilon$ beyond 0.950, again $\sigma$ decreases to zero showing that the states correspond to a synchronized region.  Thus we can observe the recurrence of synchronized state for lower values of nonisochronicity parameter $c$.  We can also observe the same behavior for higher values of nonisochronicity parameter.  We illustrate this fact for $c=5.0$ and $c=7.0$ in Figs. \ref{fg1} (b) and (c), respectively.  This also confirms the existence of recurrence of synchronized states for higher values of the nonisochronicity parameter as well. \\
{\bf(b)~~Symmetry broken global coupling:}\\
 To illustrate the dynamical behavior of the system (\ref{gs}), we choose the nonisochronicity parameter as $c=2.5$ which results in the dynamical states as depicted in Fig. \ref{fg2}(a).  For the values $\epsilon < 0.27$ the oscillators are in a synchronized state.  By increasing the strength of coupling interaction ($0.28 < \epsilon < 0.94$), $\sigma$ decreases to zero so that the system of oscillators becomes synchronized.  For the values of $\epsilon$ in the region $0.95 < \epsilon < 2.48$, $\sigma$ takes nonzero values implying that the oscillators in the system have again become desynchronized.  By increasing $\epsilon$ beyond 2.48, we can observe synchronized states where $\sigma$ takes the value zero.  Thus we can observe the recurrence of synchronized states for smaller values of nonisochronicity parameter $c$.  We also confirm the existence of recurrence of synchronized states for higher values of nonisochronicity parameter $c=4.0$ and $c=7.0$ which is illustrated in Figs. \ref{fg2} (b) and (c), respectively.  Thus we can conclude that the system of oscillators that are coupled under symmetry broken coupling follows the same transition as that of the system of oscillators with symmetry preserved coupling on the basis of the measure standard deviation $\sigma$.
\par  However, we also note an important difference in the distribution of $\sigma$ as a function of $\epsilon$ between the systems (\ref{g}) and (\ref{gs}).  While $\sigma$ varies smoothly as a function of the coupling strength $\epsilon$ (Figs. \ref{fg1}(a-c)) in the case of symmetry preserving global coupling corresponding to (\ref{fg1}), see also the work of Daido and Nakanishi \cite{6}, it varies quite randomly with symmetry broken case (\ref{gs}) as depicted in Figs. \ref{fg2}(a-c).  Thus we conclude that the presence of symmetry breaking in the system leads to an increase of disorder in the dynamical states.
\begin{figure*}[ht!]
\begin{center}
\includegraphics[width=0.8\linewidth]{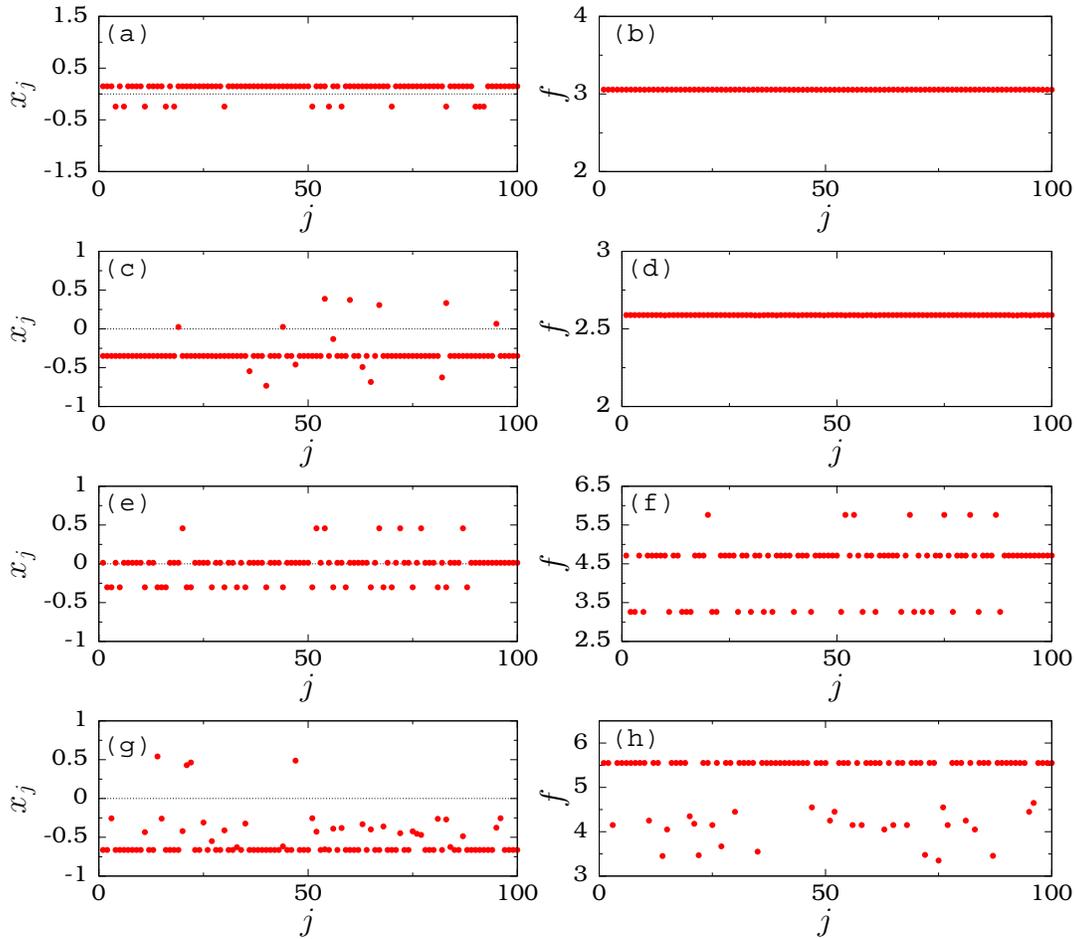}
\end{center}
\caption{Snapshots for the variables $x_j$ of the system (\ref{g}) (with symmetry preserving coupling) and the corresponding frequencies $f$ for various dynamical states.  (a)-(b) Amplitude cluster state for $c=5.0$ and $\epsilon=0.60$.  (c)-(d) Amplitude chimera state for $c=5.0$ and $\epsilon=0.75$.  (e)-(f) Frequency cluster state for $c=7.0$ and $\epsilon=0.50$.  (g)-(h) Frequency chimera states for $c=7.0$ and $\epsilon=1.0$.}
\label{fg14}
\end{figure*} 
\begin{figure}[ht!]
\begin{center}
\includegraphics[width=1.0\linewidth]{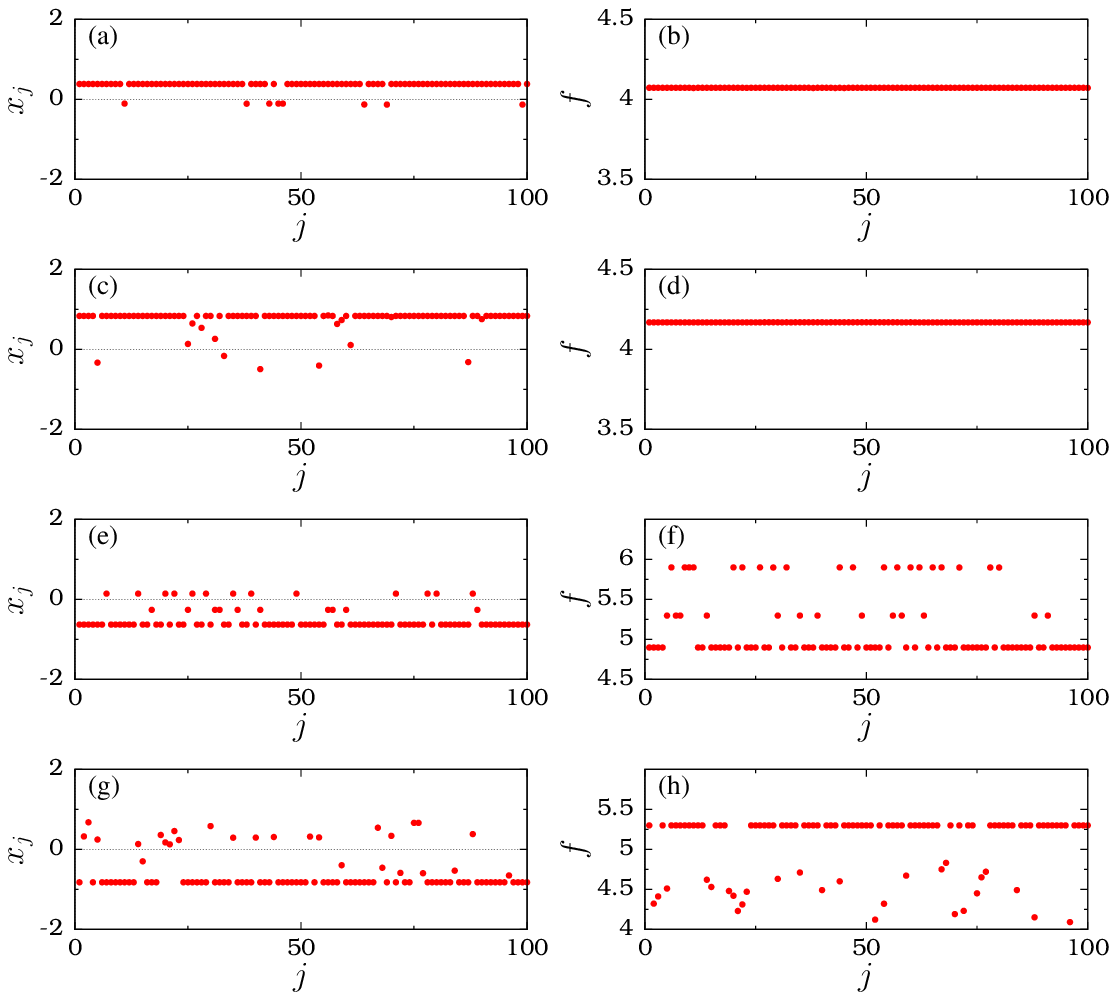}
\end{center}
\caption{Snapshots for the variables $x_j$ of the system (\ref{gs}) (with symmetry broken coupling) and the corresponding frequencies $f$ for various dynamical states.  (a)-(b) Amplitude cluster state for $c=4.0$ and $\epsilon=1.0$.  (c)-(d) Amplitude chimera state for $c=4.0$ and $\epsilon=1.30$.  (e)-(f) Frequency cluster state for $c=7.0$ and $\epsilon=0.6$.  (g)-(h) Frequency chimera state for $c=7.0$ and $\epsilon=1.0$.}
\label{fg14a}
\end{figure}
\subsubsection{CHARACTERIZATION WITH RESPECT TO STRENGTH OF INCOHERENCE}
\par By using the above notion of standard deviation introduced in \cite{6}, one cannot differentiate the chimera/cluster states and desynchronized states.  In order to know the nature of dynamical states in more detail, we look at the strength of incoherence of the system a notion introduced recently by Gopal, Venkatesan and two of the present authors \cite{39} that will help us to detect interesting collective dynamical states such as the chimera state.  For this purpose we introduce a transformation $z_j=x_j-x_{j+1}$ \cite{39}, where $j=1,2,3,...,N$.  We divide the oscillators into $M$ bins of equal length $n=N/M$ and the local standard deviation $\sigma(m)$ is defined as   
\begin{equation} 
\sigma(m)=\langle(\overline{ \frac{1}{n}\sum_{j=n(m-1)+1}^{mn} \vert z_j-\overline{z_j}\vert^2})^{1/2}\rangle_t, m=1,2,...M.
\label{sig}
\end{equation}
\par From this we can find the local standard deviation for every $M$ bin of oscillators that helps to find the strength of incoherence \cite{39} through the expression
\begin{equation} 
S=1-\frac{\sum_{m=1}^{M} s_m}{M},s_m=\Theta(\delta- \sigma(m)),
\label{soi}
\end{equation}
where $\delta$ is the threshold value which is small and $\Theta$ is the Heaviside step function. When $\sigma(m)$ is less than $\delta$, $s_m$ takes the value $1$, otherwise it is $0$. Thus the strength of incoherence measures the amount of spatial incoherence present in the system which is zero for the spatially coherent  synchronized state.  It has the maximum value, that is $S=1$, for the completely incoherent desynchronized state and has intermediate values between 0 and 1 for chimera states and cluster states.  Further to distinguish the chimera and cluster states, we make use of the quantitative measure $S_0$ \cite{40} which is the strength of incoherence before the removal of discontinuity points while $S$ is calculated in this region after removal of such points.  Here the value of $S_0$ is the same as $S$ for desynchronized, synchronized and chimera states, but for the cluster states $S_0$ takes nonzero values between zero and one while $S$ takes the value zero.  For more details see Ref. \cite{39}
\begin{figure*}[ht!]
\begin{center}
\includegraphics[width=0.9\linewidth]{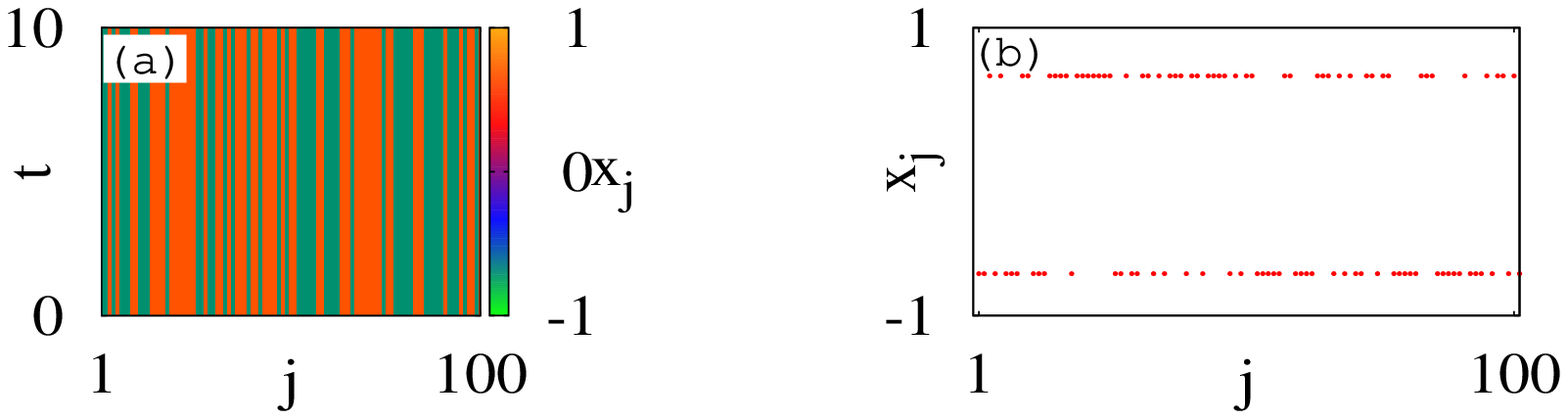}
\end{center}
\caption{(a)Space-time plot and (b) Snapshot of the variables $x_j$ for $N=100$ oscillators which confirm the existence of chimera death state for coupling strength $\epsilon=3.5$ and nonisochronicity parameter $c=2.5$.}
\label{fg6a}
\end{figure*} 
\par In order to understand the different dynamical states and their transitions more clearly, we analyze the strength of incoherence corresponding to the cases discussed in terms of $\sigma$ in the above sub-section for the symmetry preserving case (Figs. \ref{fg1}(d-f)) and symmetry breaking case (Figs. \ref{fg2}(d-f)). \\
{\bf(a)~~ Symmetry preserving global coupling:}\\
 Fig. \ref{fg1}(d) is plotted for the strength of incoherence ($S$ as well as $S_0$) with respect to the coupling strength $\epsilon$ for $c=2.5$.  Initially the oscillators are desynchronized in the region $\epsilon< 0.27$ where $S$ (and also $S_0$) takes the value unity.  By increasing the coupling strength $\epsilon$, $S$ (and also $S_0$) reaches the value zero where the oscillators are in synchronization (region-I).  Further for the values of $\epsilon$ between $0.675$ and $0.98$, $S_0$ takes nonzero value (below one) while $S$ takes the value zero (region-II) indicating the presence of cluster states.  Interestingly, on further increasing $\epsilon$ again we can observe the synchronized state where the value of ($S,S_0$) is zero (region-I$^\prime$).  Thus we can conclude that the synchronized state is mediated through cluster states for lower values of the nonisochronicity parameter $c$. 
\par Further we address the question whether the synchronized state is mediated through cluster states for higher values of nonisochronicity parameters also.  This is analyzed by the behavior of $S$ and $S_0$ for the values $c=4.0$ and $c= 7.0$ which is depicted in Figs. \ref{fg1}(e) and \ref{fg1}(f), respectively.  When $c=4.0$ (Fig. \ref{fg1}(e)), we can observe that for small values of $\epsilon$ ($\epsilon< 0.27$) $S$ and $S_0$ take the value unity which shows the desynchronization of oscillators in this region and by increasing $\epsilon$ to $0.28$, the oscillators are synchronized where the value of $S$ and $S_0$ decreases to zero.  We can observe the presence of cluster states (where $S=0$ and $0<S_0<1$, region-II) in the range $0.96<\epsilon<1.23$.  Interestingly, further increase in $\epsilon$ leads to the occurrence of chimera states where $S$ (and also $S_0$) oscillates between zero and one (region-III).  Again beyond $\epsilon=1.23$ we can observe that the value of $S=0$ and $S_0$ takes the value between zero and one confirming the presence of cluster states (region-II$^\prime$).  Further increase of $\epsilon$ leads to synchronization of oscillators with $S=0$ (and also $S_0=0$) in the region-I$^\prime$.  
\par Next, a study of the strength of incoherence for $c=7.0$ confirms a similar transition behavior which is depicted in Fig. \ref{fg1}(f).  Hence we conclude that the synchronized state is mediated through chimera states in addition to the cluster states for higher values of nonisochronicity parameter $c$. \\
{\bf(b)~~Symmetry broken global coupling:}\\
Next, Figs. \ref{fg2}(d-f) depict the behavior of $S$ (and $S_0$) for the system (\ref{gs}) with symmetry breaking in the coupling for three values of the nonisochronicity parameter.  For $c=3.0$, the oscillators follow the transition route (Fig. \ref{fg2}(d)) as desynchronization $\rightarrow$ synchronization $\rightarrow$ cluster states $\rightarrow$ synchronization.  When $c=4.0$ (Fig. \ref{fg2}(e)) and $c=7.0$ (Fig. \ref{fg2}(f)) the transition route is represented as desynchronization $\rightarrow$ synchronization $\rightarrow$ cluster states $\rightarrow$ chimera states $\rightarrow$ cluster states $\rightarrow$ synchronization.  We also note here that the region of chimera states is much wider in the present case, Figs. \ref{fg2}(d-f), compared to the symmetry preserving case as seen in Figs. \ref{fg1}(d-f).  Hence the characterization with respect to $S$ also confirms that the symmetry breaking in the coupling leads to an increase in disorder in the dynamical states and also the regions are widened, while both the system (\ref{g}) and (\ref{gs}) follow the same transition route.  In addition symmetry breaking in the system (\ref{gs}) ultimately leads to a specific feature called chimera death which is demonstrated in the following sub-section.
\begin{figure}[ht!]
\begin{center}
\includegraphics[width=1.0\linewidth]{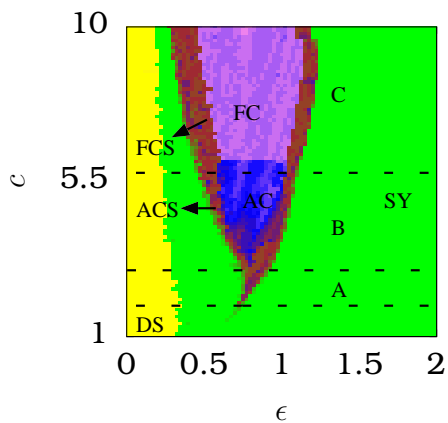}
\end{center}
\caption{Phase diagram for the system (\ref{g}) (symmetry preserved coupling).  Green color shows the region of synchronized states, yellow color shows desynchronized state (DS) region, blue color represents amplitude chimera state (AC), violet color shows the frequency chimera state (FC) and brown color represents the amplitude cluster state (ACS) and frequency chimera state (FCS). Region `A': DS $\rightarrow$ACS$\rightarrow$ SY, region `B': DS $\rightarrow$ACS$\rightarrow$ AC $\rightarrow$ACS $\rightarrow$ SY, and region `C' : DS $\rightarrow$FCS$\rightarrow$ FC $\rightarrow$FCS $\rightarrow$ SY.}
\label{fga}
\end{figure}  
\begin{figure*}[ht!]
\begin{center}
\includegraphics[width=0.8\linewidth]{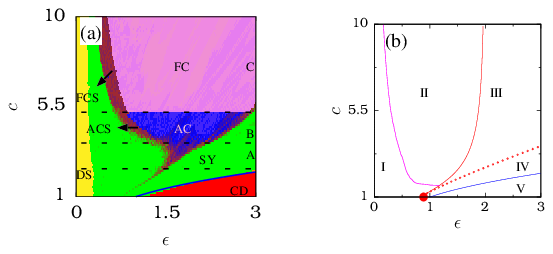}
\end{center}
\caption{(a) Phase diagram for the globally coupled system (\ref{gs}) with symmetry breaking by varying the values of $c$ and  $\epsilon$ for $N=100$ oscillators.  Green color shows the synchronized state, yellow color shows the state corresponding to desynchronized region (DS), blue color shows amplitude chimera state (AC), violet color shows the frequency chimera state (FC) and brown color shows the amplitude cluster states (ACS) and frequency chimera states (FCS) and red color shows the chimera death region.   Region `A': DS $\rightarrow$ACS$\rightarrow$ SY $\rightarrow$ CD, region `B': DS $\rightarrow$ACS$\rightarrow$ AC $\rightarrow$ACS $\rightarrow$ SY $\rightarrow$ CD, and region `C' : DS $\rightarrow$FCS$\rightarrow$ FC $\rightarrow$FCS $\rightarrow$ SY $\rightarrow$ CD. (b) shows the analytical plot for global coupling with symmetry breaking. Regions ($I$) and ($IV$) correspond to synchronized states, region ($II$) and ($III$) are for chimera and cluster states and the region ($V$) shows the chimera death states.  The red color dot represents the Takens-Bogdanov point.}
\label{fg10}
\end{figure*} 
\subsubsection{CHARACTERIZATION OF CLUSTER, CHIMERA, AND CHIMERA DEATH STATES}
\par To identify the nature of the cluster and chimera states (and also the chimera death state) in more detail, we look into the snapshots for the variables $x_j$ and the corresponding frequencies of the oscillators.  The frequency of each of the oscillators is calculated from the expression $f_j= 2\pi \Omega_j/\Delta T$, where $j=1,2,3,...N$, $\Omega_j$ is the number of maxima in the time series $x_j$ of the $j^{th}$ oscillator with time interval $\Delta T$. \\
{\bf (a)~~Symmetry preserving coupling:}\\
At first we consider the system (\ref{g}) for the three different values of the nonisochronicity parameter $c$ (similar to the above discussed cases) which are illustrated in Figs. \ref{fg14}.  Snapshots of the variables $x_j$ and their corresponding frequencies $f_j$ (Fig. \ref{fg14}(a) and (b)) are plotted for  $c=5.0$ and $\epsilon=0.60$ confirming the presence of cluster states.  In this case, the oscillators within the each distinct cluster have identical amplitudes and phases.  However the frequencies of the oscillators in different groups are the same [Fig. \ref{fg14} (b)].  This type of cluster states is designated as {\it amplitude cluster states}.  By increasing $\epsilon$ to 0.75, we can also observe the {\it amplitude chimera state} (the fluctuations exist only in the amplitudes while the frequencies of all the oscillators are the same) which is shown in Figs. \ref{fg14}(c) and (d). 
\par  Interestingly, a further increase in the nonisochronicity parameter leads to an increase in disorder in the frequency of the dynamical states as well.  For $c=7.0$, we can observe the frequency cluster states.  This means that the oscillators within each cluster have identical amplitudes and phases and also the same frequencies. These are different for different groups as shown in Figs. \ref{fg14}(e,f) for $\epsilon=0.50$. This type of cluster states are designated as frequency cluster states. Figs. \ref{fg14}(g) and (h) are the snapshot and frequencies of the oscillators corresponding to a frequency chimera state for $\epsilon=1.0$.  Here the frequencies of the oscillators in the coherent region are the same, while they are randomly distributed in the region corresponding to incoherent behavior [Fig. \ref{fg14}(h)].  Apart from the above described states, no other dynamical state (chimera death state) has been identified in this case of symmetry preserving coupling. \\
{\bf (b)~~Symmetry broken coupling:}\\
Next we discuss the nature of cluster and chimera states for the system (\ref{gs}) in the case of symmetry broken coupling.  The snapshots for the variables $x_j$ and the frequencies of the oscillators which are shown in Figs. \ref{fg14a}(a) and (b) respectively show the presence of amplitude cluster states for $c=4.0$ and $\epsilon=1.0$.  On further increasing $\epsilon$ to $\epsilon=1.30$, one finds the existence of amplitude chimera states (Figs. \ref{fg14a}(c), (d)).  For higher values of the nonisochronicity parameter ($c=7.0$), we can observe the frequency cluster states for $\epsilon=0.6$ (Figs. \ref{fg14a}(e), (f)) and frequency chimera states for $\epsilon=1.0$ (Figs. \ref{fg14a}(g), (h)).  
\par Finally, while varying the coupling strength, the symmetry breaking present in the coupling leads to  {\it chimera death state} for larger values of $\epsilon$.  It has the combined properties of chimera and oscillation death. The population of identical oscillators splits into two coexisting domains: (i) spatially coherent oscillation death (neighboring oscillators populate in the same branch of inhomogeneous steady state either as $x^{(1)}$ or $x^{(2)})$ and (ii) spatially incoherent oscillation death (population of neighboring oscillators are completely random between $x^{(1)}$ and $x^{(2)}$) as shown in Fig. \ref{fg6a} for $c=2.5$ and $\epsilon=3.5$, where $x^{(1)}$ and $x^{(2)}$ represent the amplitudes of the oscillators corresponding to the upper and lower branches of the inhomogeneous steady state, respectively.  Here the total number of oscillators (100) splits into two equal groups of inhomogeneous steady state.  Finally the diverse transition routes to chimera death state is identified with a two parameter phase diagram in the next section. 
\section{DIFFERENT TRANSITIONS IN THE PARAMETRIC SPACE ($\epsilon,c$)} 
\par We now present a comprehensive analysis of the different dynamical states and transition routes between them in the ($\epsilon,c$) two parameter space under both symmetry preserved and symmetry broken couplings.
\subsection{ Global coupling under symmetry preservation}
\par In order to study the appearance of the swing of synchronized state in globally coupled oscillators intensely, we plotted the two parameter phase diagram in the parametric space ($\epsilon,c$) in terms of the strength of incoherence $S$, as shown in Fig. \ref{fga}.  Different dynamical states for a given pair of values of  $\epsilon$ and $c$ are identified by making  use of the strength of incoherence ($S$).  For this purpose we scan the ( $\epsilon,c$) plane with increments of 0.015 in  $\epsilon$ and 0.045 in $c$.  We choose random initial conditions between -1 to +1 for every choice of  $\epsilon$ and $c$ values and observe the dynamics in the two parameter space in terms of $S$.  For sufficiently small values of $c$, the system of oscillators attains a synchronized state directly from the desynchronized state.  By increasing $c$, for $c=2.5$ the synchronized state is mediated through amplitude cluster states and it follows the transition route as {\textit{desynchronization $\rightarrow$ synchronization $\rightarrow$ amplitude cluster states $\rightarrow$ synchronization}.  This route is illustrated in the region `A'.  We can observe that the synchronized  state is mediated through amplitude chimera states along with amplitude cluster states in the range $3.0 \leq c\leq 6.0$ and the corresponding transition route is represented in the region `B'.  This transition is the one reported by Daido and Nakanishi in \cite{6}.  They also identified that the cluster states exist as periodic and nonperiodic (quasi-periodic) desynchronized states.  In addition we distinguish the cluster states as amplitude cluster states and amplitude chimera states and the transition route is represented as {\textit{desynchronization $\rightarrow$ synchronization $\rightarrow$ amplitude cluster states $\rightarrow$ amplitude chimera  $\rightarrow$ amplitude cluster states$ \rightarrow$ synchronization}.  Interestingly, on further increasing the value of $c$ to 7.0, the synchronized states of the oscillators are mediated through the frequency chimera states in addition to the frequency cluster states which is represented by the region `C' as {\textit{desynchronization $\rightarrow$ synchronization $\rightarrow$ frequency cluster states $\rightarrow$ frequency chimera  $\rightarrow$ frequency cluster states$ \rightarrow$ synchronization}.  A summary of different transition routes is presented in Table I.
\subsection{ Global coupling under symmetry breaking}
\par Next, for the case of globally coupled oscillators with symmetry broken coupling over a wide range of coupling strength, we plot the two parameter phase diagram in the parametric space ($\epsilon,c$) in terms of the strength of incoherence $S$ which is shown in Fig. \ref{fg10} (a).  One of the major distinguishing features in the present case is that the chimera death states occur for all values of the nonisochronicity parameter $c$ for suitable values of $\epsilon$.  Further the range of chimera states gets widened as a function of $\epsilon$.  Specifically we can observe that for sufficiently small values of $c$ ($c \leq 1.5$) the system of oscillators attains chimera death state through a synchronized state from the desynchronized state.  By increasing $c$ to $c=2.5$, the synchronized state gets mediated through the amplitude cluster state and attains the chimera death state.  This route is represented in the region `A' (Fig.\ref{fg10} (a)) and it follows the transition route as \textit{ desynchronization $\rightarrow$ synchronization $\rightarrow$ amplitude cluster $\rightarrow$ synchronization $\rightarrow$ chimera death}.   
On increasing the value of the nonisochronicity parameter to the range $2.5 \leq c\leq 5$, we can observe the swing of synchronized state via amplitude chimera state (the oscillators in the coherent and incoherent regions having the same frequency) in addition to the  amplitude cluster state, but ultimately ending up in the chimera death state (beyond the region of $\epsilon$ shown in Fig. 7 (a)).  We can observe the appearance of amplitude chimera states in between the regions of the amplitude cluster states.  Here it follows the transition route as \textit{desynchronization $\rightarrow$ synchronization $\rightarrow$ amplitude cluster $\rightarrow$ amplitude chimera  $\rightarrow$ amplitude cluster$ \rightarrow$ synchronization $\rightarrow$ chimera death} (region `B' in Fig. (\ref{fg10} (a))).  Interestingly, on further increasing $c$ beyond $c\approx 5$, the nonisochronicity parameter induces a disorder in the frequencies of the dynamical states and causes the system to pass through a different set of dynamical states.  It follows the transition route as \textit {desynchronization $\rightarrow$ synchronization $\rightarrow$ frequency cluster $\rightarrow$ frequency chimera  $\rightarrow$ frequency cluster$\rightarrow$ synchronization $\rightarrow$ chimera death} which is shown in the region `C'.  
\par In our study, we analyzed the dynamical system (\ref{gs}) over a wide range of nonisochronicity parameter and the main points are as follows.  We identified the presence of amplitude chimera states.  We have also observed that the synchronized states are mediated through the amplitude chimera states in addition to the cluster states in (\ref{gs}).  Further we identified that the chimera regions get widened due to the presence of symmetry breaking in the coupling.  The chimera death states occur for sufficiently large $\epsilon$ values for all values of $c$.   We have also characterized the cluster states which appeared in the regions `A', `B' and `C' as amplitude and frequency clusters.  The various transition routes are again summarized in Table I.
\subsection{Analytical results}
\par To understand the existence of the various states discussed earlier analytically, we first consider the system (\ref{gs}) and then (\ref{g}).\\
{\bf (i)~~Symmetry broken case:}\\
~~We start by assuming the chimera states correspond to the coexistence of synchronized and desynchronized identical groups of oscillators so that the system (\ref{gs}) can be written as 
 \begin{eqnarray}
\frac{d z_s}{dt}=z_s-(1-ic)|z_s|^2z_s+ \qquad \qquad \qquad \nonumber\\
\qquad \epsilon(pRe(Z_s)+(1-p)Re(Z_d)-Re(z_s)), \label{sh1}\\
\frac{d z_d}{dt}=z_d-(1-ic)|z_d|^2z_d+\qquad \qquad \qquad\nonumber\\
\epsilon(pRe(Z_s)+(1-p)Re(Z_d)-Re(z_d)),\label{sh2}
\end{eqnarray}
where $s=1,2,3...l$, $d=1,2,3,...k$, $l + k =N$, $N$ being the number of oscillators, $p=l/N$, $q=k/N$, $p+q=1$, $Z_s=(1/l)\sum_{s=1}^{l}z_s$, $Z_d=(1/k)\sum_{k=1}^{k}z_d$.  $z_{s}$ and $z_{d}$ are the states of the oscillators corresponding to synchronized and desynchronized states, respectively.  Following refs. \cite{6, 40, 41}, considering the $p \approx 1$ limit, we can see that the chimera state appears from the synchronized state by varying the coupling strength  and we obtain $z_s=e^{ict}$ and 
\begin{eqnarray}
\frac{d z_d}{dt}&=&z_d-(1-ic)|z_d|^2z_d-\epsilon(Re(e^{ict})-Re(z_d)).\nonumber
\end{eqnarray}
The above equation has a  solution $z_d=e^{ict}$ that implies the completely synchronized manifold of the system which is always stable.  In order to find the stability of the chimera/cluster state we apply a multi-time scale perturbation to the solution of $z_d$ so that $z_d=w(\tau)e^{ic t}$, where $w(\tau)$ represents the amplitude of the desynchronized oscillators and $t=t_0+\tau$ with $t_0$ and $\tau$ representing the fast and slow time scales, respectively. Thus the dynamics of each of the desynchronized oscillators is represented by
\begin{eqnarray}
\frac{dw}{d\tau}=(1-ic)w-(1-ic)|w|^2w+\frac{\epsilon}{2}(1-w)+\nonumber\\
\frac{\epsilon}{2}(1-{\bar{w}}))e^{-2ic t}.\quad \quad
\end{eqnarray}
Averaging over the fast time scale $t_0$ between $0$ and $\frac{2\pi}{c}$ we obtain
\begin{eqnarray}
\frac{d w}{d\tau}&=&(1-ic)w-(1-ic)|w|^2w+\frac{\epsilon}{2}(1-w).
\label{wd}
\end{eqnarray}
Eq. (\ref{wd}) has one stable fixed point at $w=1$, which corresponds to complete synchronization manifold. The other two fixed points
\begin{small}
\begin{eqnarray}
|w_{2,3}|^2=-\frac{\epsilon -(1+c^2) \pm \sqrt{-\left(1+c^2\right) \epsilon ^2+\left(1+c^2-\epsilon \right)^2}}{2 \left(1+c^2\right)}\quad
\label{fix}
\end{eqnarray}
\end{small}
exist for 
\begin{eqnarray}
\epsilon<\epsilon_{SN}=-\frac{1+c^2-\sqrt{\left(1+c^2\right)^3}}{c^2}.
\label{sn}
\end{eqnarray}
From the linear stability analysis on equation (\ref{wd})  about the fixed points (\ref{fix}) we get the Hopf bifurcation curve,
\begin{eqnarray}
\epsilon_{H}=\frac{2 \left(-2+\sqrt{5+2 c^2+c^4}\right)}{1+c^2}.
\label{h}
\end{eqnarray}
\par The curves for saddle node and Hopf bifurcations are plotted by using Eqs. (\ref{sn}) and (\ref{h}) which are shown by the red color solid and red color dotted lines in Fig. \ref{fg10}(b).  Saddle connection boundary is obtained by solving (\ref{wd}) numerically and its boundary is given by the pink color line.  Takens-Bogdanov point is denoted by $\epsilon_{TB}=2 \left(-1+\sqrt{2}\right)$ at $|c|=1$.  Thus, it turns out that for $|c|>1$, the original system has a stable chimera/cluster state solution in the range $\epsilon_{SC} < \epsilon < \epsilon_{SN}$.  In Fig. \ref{fg10}(b) regions $I$ and $IV$ correspond to synchronized states, regions $II$ and $III$ represent cluster and chimera states and the region $V$ corresponds to the chimera death states as discussed below.
\par On the other hand, chimera death represents the situation where the total population is split into two groups of inhomogeneous steady states.  Eqs (\ref{sh1}) and  (\ref{sh2}) can be written as two populations of oscillators and is specified by
\par \begin{eqnarray}
\frac{d z_{h1}}{dt}=z_{h1}-(1-ic)|z_{h1}|^2z_{h1}+\qquad \qquad \qquad\nonumber\\
\epsilon(pRe(z_{h1})+(1-p)Re(z_{h2})-Re(z_{h1})), \label{de}\\
\frac{d z_{h2}}{dt}=z_{h2}-(1-ic)|z_{h2}|^2z_{h2}+\qquad \qquad \qquad\nonumber\\
\epsilon(pRe(z_{h1})+(1-p)Re(z_{h2})-Re(z_{h2})),
\label{de1}
\end{eqnarray}
where $z_{h1}$ and $z_{h2}$ are the states of the oscillators corresponding to two groups of inhomogeneous steady states.  Considering our numerical results, the above system has a trivial equilibrium point $(x_1,y_1,x_2,y_2)$=(0,0,0,0) and a nontrivial (inhomogeneous) equilibrium point $(x_1,y_1,x_2,y_2)$=$(x_1^*,y_1^*,-x_1^*,-y_1^*)$ for $p=0.5$ (as confirmed numerically), where 
\begin{eqnarray}
x_1^*= \frac{2 \sqrt{2}c y_1^*}{-\epsilon +\alpha},
\label{f1}
\end{eqnarray}
\begin{eqnarray}
y_1^*=\frac{(-\epsilon +\alpha)\sqrt{-\frac{(-1+\epsilon)(\epsilon+\alpha)+c^2(-4+3\epsilon+\alpha)}{(1-c)^2 \epsilon}}}{2\sqrt{2}c},
\label{f2}
\end{eqnarray}  with $\alpha=\sqrt{4c^2(-1+\epsilon)+\epsilon^2}$.  We find the above nontrivial fixed point is linearly stable for $\epsilon \geq \epsilon_c$, where the critical value of $\epsilon$ is given by  
\begin{eqnarray}
\epsilon_c=\frac{1}{2}(1+c)^2. 
\label{de3}
\end{eqnarray}
By using the above equation, we plotted the curve for the chimera death region (blue color line in Fig. \ref{fg10}(a) and (b)) for two populations of oscillators.  Fitting this curve with numerically plotted phase diagram, we can observe that analytical results closely match with the numerical results (see Fig.  \ref{fg10}(a)).\\
{\bf (ii)~~Symmetry preserving case:}
\par Proceeding in a similar way for the symmetry preserved global coupling system (\ref{g}), Eqs. (\ref{sh1}) and (\ref{sh2}) can be rewritten as
 \begin{eqnarray}
\frac{d z_s}{dt}=z_s-(1-ic)|z_s|^2z_s+ \qquad \qquad \qquad \nonumber\\
\qquad \epsilon(p z_s+(1-p)z_d-z_s),\label{sh11}\\
\frac{d z_d}{dt}=z_d-(1-ic)|z_d|^2z_d+\qquad \qquad \qquad\nonumber\\
\epsilon(p z_s+(1-p)z_d-z_d).
\label{sh23}
\end{eqnarray}
As before, the slow scale dynamics of the each of the desynchronized oscillators can be represented as \cite{6} 
\begin{eqnarray}
\frac{d w}{d\tau}&=&(1-ic)w-(1-ic)|w|^2w+\epsilon(1-w).
\label{wdw}
\end{eqnarray}
By replacing $\epsilon$ by $\epsilon/2$ in the above equation, it can be reduced to Eq. (\ref{wd}) which corresponds to the case of symmetry breaking in the coupling.  This implies that the dynamical regions in the system (\ref{gs}) get widened approximately twice as that of the system (\ref{g}).  From (\ref{sh11}) and (\ref{sh23}), looking for chimera death states, the two population equations (\ref{de}) and (\ref{de1}) can now be written as  
\par \begin{eqnarray}
\frac{d z_{h1}}{dt}=z_{h1}-(1-ic)|z_{h1}|^2z_{h1}+\qquad \qquad \qquad\nonumber\\
\epsilon(pz_{h1}+(1-p)z_{h2}-z_{h1}),\label{hde}\\
\frac{d z_{h2}}{dt}=z_{h2}-(1-ic)|z_{h2}|^2z_{h2}+\qquad \qquad \qquad\nonumber\\
\epsilon(pz_{h1}+(1-p)z_{h2}-z_{h2}).
\label{hde1}
\end{eqnarray}
We can check that the fixed points of the type (\ref{f1})-(\ref{f2}) do not exist for Eqs. (\ref{hde}) and (\ref{hde1}) under symmetry preserved coupling and so one cannot identify the existence of chimera death in the system (\ref{g}).
\section{conclusion}
\par In summary, we have investigated the common and distinguishing features underlying the collective dynamics of globally coupled Stuart-Landau oscillators under two different coupling schemes: (1). symmetry preserved coupling, (2). symmetry broken coupling.  We have observed that the synchronized state is mediated through the amplitude chimera states in addition to amplitude cluster states for lower values of the nonisochronicity parameter.  For higher values of this parameter, the synchronized state gets mediated through the frequency chimera states in addition to frequency cluster states.  Moreover, the presence of symmetry breaking in the coupling leads to increased disorder in the dynamical states and also leads to the widening of the various interesting dynamical regions.  In addition we have also identified the existence of chimera death states and diverse transitions routes to chimera death state in the case of symmetry broken coupling. 
\par Through a multi-time scale perturbation analysis, we have also analytically established the various regions of dynamical states including chimeras and chimera death states.  We can thus conclude that symmetry breaking in global coupling leads to  a rich variety of collective dynamical states.
\section*{Acknowledgements}
The work of  KP and MS forms part of a research project sponsored by NBHM, Government of India.  The work of VKC forms part of a research project sponsored by INSA Young Scientist Project and while the work of ML forms part of an IRHPA project.  ML also acknowledges the financial support under a DAE Raja Ramanna Fellowship. 
\appendix
\section{Transition routes between various collective dynamical states}
\par The different transition routes allowed by the coupled Stuart-Landau oscillators under the two different coupling schemes are summarized in table $I$ for comparison purpose. 
\begin{table*}[]
\begin{center}
\begin{tabular}{|c|c|c|}
\hline
{\bf S. No} & {\bf Coupling schemes} & {\bf Transition routes} \\
\hline	
1 &global coupling & (i)~~DS $\rightarrow$ SY ~~~~~~~~~~~~~~~~~~~~~~~~~~~~~~~~~~~~~~~~~~~~~~~~~ \\
&~~without symmetry breaking &  (ii)~~DS $\rightarrow$ACS$\rightarrow$ SY ~~~~~~~~~~~~~~~~~~~~~~~~~~~~~~~~~~~~~~~~\\
& &(iii)~~DS $\rightarrow$ACS$\rightarrow$ AC $\rightarrow$ACS $\rightarrow$ SY~~~~~~~~~~~~~~~~~~~~~ \\
& &(iv)~~DS $\rightarrow$FCS $\rightarrow$ FC$\rightarrow$ FCS $\rightarrow$ SY~~~~~~~~~~~~~~~~~~~~~\\
\hline
2 &global coupling & (i)~~DS $\rightarrow$ SY $\rightarrow$ CD~~~~~~~~~~~~~~~~~~~~~~~~~~~~~~~~~~~~~~~~ \\
&~~with symmetry breaking &  (ii)~~DS $\rightarrow$ACS$\rightarrow$ SY $\rightarrow$ CD~~~~~~~~~~~~~~~~~~~~~~~~~~~~~~~~~\\
& &(iii)~~DS $\rightarrow$ACS$\rightarrow$ AC $\rightarrow$ACS $\rightarrow$ SY$\rightarrow$ CD ~~~~~~~~~~~~ \\
& &(iv)~~DS $\rightarrow$FCS $\rightarrow$ FC$\rightarrow$ FCS $\rightarrow$ SY $\rightarrow$ CD~~~~~~~~~~~~\\
\hline
\end{tabular}
\caption{Different coupling schemes and corresponding transition routes.  DS $\rightarrow$ Desynchronized states, SY$\rightarrow$ synchronized states, AC$\rightarrow$ amplitude chimera states, FC $\rightarrow$ frequency chimera states, CD $\rightarrow$ chimera death states, ACS $\rightarrow$ amplitude cluster states, FCS $\rightarrow$ frequency cluster states.}
\end{center}
\end{table*}

\end{document}